\documentclass[lettersize,journal]{IEEEtran}
\usepackage{amsmath,amsfonts}
\usepackage{array}
\usepackage[caption=false,font=normalsize,labelfont=sf,textfont=sf]{subfig}
\usepackage{textcomp}
\usepackage{stfloats}
\usepackage{url}
\usepackage{verbatim}
\usepackage{graphicx}
\usepackage{cite}
\usepackage{multirow}
\usepackage{float}
\usepackage{amsfonts,amssymb}
\usepackage{amsthm}
\usepackage{tabularx}
\usepackage{bbding}
\usepackage{color,xcolor}
\usepackage{booktabs}
\usepackage{hyperref}
\usepackage{verbatim}
\usepackage[table]{xcolor}
\usepackage[linesnumbered,lined,boxed,commentsnumbered,ruled]{algorithm2e}
\makeatletter
\renewcommand{\maketag@@@}[1]{\hbox{\m@th\normalsize\normalfont#1}}
\makeatother
\begin{document}

\title{IML-Spikeformer: Input-aware Multi-Level Spiking Transformer\\ for Speech Processing}

\author{Zeyang Song, Shimin Zhang, Yuhong Chou, Jibin Wu$^\dagger$,~\IEEEmembership{Member,~IEEE}, Haizhou Li$^\dagger$,~\IEEEmembership{Fellow,~IEEE}
\thanks{
The work is partly supported by the National Natural Science Foundation of China (Grant No. 62271432 and 62306259),  Shenzhen Science and Technology Program (Shenzhen Key Laboratory, Grant No. ZDSYS20230626091302006), and the Program for Guangdong Introducing Innovative and Entrepreneurial Teams, Grant No. 2023ZT10X044, and the Research Grants Council of the Hong Kong SAR (Grant No. C5052-23G, PolyU15217424, PolyU25216423).

Zeyang Song and Haizhou Li are with the Department of Electrical and Computer Engineering,
National University of Singapore, Singapore 119077

Haizhou Li is also with the School of Artificial Intelligence, The Chinese University of Hong Kong, Shenzhen, 518172 China; Shenzhen Loop Area Institute, Shenzhen, China.

Shimin Zhang, Yuhong Chou and Jibin Wu are
with the Department of Data Science and Artificial Intelligence, The Hong Kong Polytechnic University, Hong Kong SAR. Jibin Wu is also with the
Department of Computing, The Hong Kong Polytechnic University, Hong Kong SAR($^\dagger$Corresponding Author: jibin.wu@polyu.edu.hk; haizhouli@cuhk.edu.cn)
}
}
\markboth{Journal of \LaTeX\ Class Files,~Vol.~14, No.~8, August~2021}%
{Shell \MakeLowercase{\textit{et al.}}: A Sample Article Using IEEEtran.cls for IEEE Journals}


\maketitle

\begin{abstract}
Spiking Neural Networks (SNNs), inspired by biological neural mechanisms, represent a promising neuromorphic computing paradigm that offers energy-efficient alternatives to traditional Artificial Neural Networks (ANNs). Despite proven effectiveness, SNN architectures have struggled to achieve competitive performance on large-scale speech processing tasks. Two key challenges hinder progress: (1) the high computational overhead during training caused by multi-timestep spike firing, and (2) the absence of large-scale SNN architectures tailored to speech processing tasks. To overcome the issues, we introduce Input-aware Multi-Level Spikeformer, i.e. IML-Spikeformer, a spiking Transformer architecture specifically designed for large-scale speech processing. Central to our design is the Input-aware Multi-Level Spike (IMLS) mechanism, which simulates multi-timestep spike firing within a single timestep using an adaptive, input-aware thresholding scheme. IML-Spikeformer further integrates a Re-parameterized Spiking Self-Attention (RepSSA) module with a Hierarchical Decay Mask (HDM), forming the HD-RepSSA module. This module enhances the precision of attention maps and enables modeling of multi-scale temporal dependencies in speech signals. Experiments demonstrate that IML-Spikeformer achieves word error rates of 6.0\% on AiShell-1 and 3.4\% on Librispeech-960, comparable to conventional ANN transformers while reducing theoretical inference energy consumption by 4.64$\times$ and 4.32$\times$ respectively. IML-Spikeformer marks an advance of scalable SNN architectures for large-scale speech processing in both task performance and energy efficiency.  Our source code and model checkpoints are publicly available at github.com/Pooookeman/IML-Spikeformer
\end{abstract}


\section{Introduction}

\IEEEPARstart{R}{ecent} advances in speech processing benefit from large-scale deep learning models, particularly Transformer-based architectures~\cite{latif2023transformers}. The high computational cost of such models has motivated the search for energy-efficient alternatives. Spiking Neural Networks (SNNs), which mimic the event-driven information processing of biological neurons, have emerged as a promising solution. Unlike traditional artificial neural networks (ANNs), SNNs emulate the dynamics of biological neurons and utilize spike trains for the representation of information~\cite{maass1997networks}. The event-driven nature of Spiking Neural Networks (SNNs) enables asynchronous computation on neuromorphic hardware~\cite{merolla2014million, davies2018loihi, pei2019towards}, where a neuron responds only upon arrival of a spike. This sparse activation mechanism significantly reduces power consumption, making SNNs particularly suitable for energy-constrained edge computing applications~\cite{duan2023neurozoom, lenk2023neuromorphic}.

Beyond their energy efficiency advantages, SNNs have demonstrated remarkable performance across diverse application domains. Substantive empirical evidence validates their effectiveness in computer vision~\cite{fang2021deep, zhou2022spikformer, zhang2023direct, yao2024spikev2}, 
 and natural language processing~\cite{zhu2023spikegpt, xing2024spikelm}. In speech processing, SNNs have achieved promising performance in keyword spotting (KWS)~\cite{cramer2020heidelberg, yang2022deep, song2024spiking}, acoustic event and sound classification~\cite{wu2018spiking, zhou2024enhancing}, and automatic speech recognition (ASR)~\cite{wu2020deep, bittar2022surrogate, ponghiran2022spiking, wang2023complex}. 

Despite promising results, SNNs are not without issues. For example, the challenges in training strategy\cite{wu2018spatio, pmlr-v202-wang23j} and architectures~\cite{fang2021deep, eshraghian2023training} prevent them from scaling up to large networks for complex real-world tasks, such as large-vocabulary ASR~\cite{wu2020deep}. There have been many attempts to address the issues. By integrating self-attention mechanisms into SNN architecture~\cite{zhou2022spikformer, yao2023spike, yao2024spikev2}, spiking Transformer models are proposed to enhances model's sequential modeling capability for long-range and complex temporal dependencies. The spiking Transformer employs multi-timestep firing to increase SNN's representational power by introducing a performance-efficiency tradeoff. 

While several methods~\cite{li2021differentiable, meng2022training, yang2023snib, yang2023effective} are proposed to reduce the information loss in spike firing~\cite{guo2022loss} and yields substantial performance gains with increase time windows, it also introduces computational complexity that scales linearly with window length, increasing both time and memory requirements~\cite{fang2023spikingjelly}. This growing computational overhead ultimately limits the scope of large-scale SNN applications~\cite{xiao2022online, yang2022training}. Moreover, current research on spiking Transformers has predominantly focused on computer vision tasks~\cite{zhou2022spikformer, yao2023spike, yao2024spikev2}, with their potential for speech processing remaining largely unexplored.

In this work, we aim to develop a spiking Transformer architecture for speech processing tasks with efficient training schemes. To mitigate the computational overhead of multi-timestep firing during training, we propose an \textit{Input-aware Multi-Level Spike (IMLS)} firing mechanism. IMLS enables the simulation of multi-timestep dynamics within a single timestep, thereby reducing training costs while retaining the spike-driven computational paradigm at inference. Moreover, IMLS introduces an input-aware adaptation strategy that adjusts neuronal firing thresholds according to the statistical distribution of pre-synaptic inputs, which improves both representational expressivity and training stability.

Building on IMLS, we design the \textit{IML-Spikeformer}, a spiking Transformer tailored for speech processing. Its central component is the \textit{Hierarchical Decay Re-parameterized Spiking Self-Attention (HD-RepSSA)} module, which enhances the representational capacity of the spiking attention map via a re-parameterization scheme while preserving spike-driven efficiency. Inspired by the auditory system’s multi-scale temporal dynamics, we further propose a \textit{Hierarchical Decay Mask (HDM)} that modulates attention maps across network layers, enabling the model to capture multi-scale temporal dependencies crucial for speech tasks.

Our main contributions are summarized as follows:
\begin{itemize}
\item \textbf{IMLS mechanism:} We introduce a multi-level spike firing strategy that reduces the training cost of multi-timestep firing, while an input-aware adaptive threshold aligned with pre-synaptic input statistics enhances model expressivity and stabilizes training.
\item \textbf{IML-Spikeformer architecture:} We propose a novel spiking Transformer featuring the HD-RepSSA module for re-parameterized spiking self-attention and the HDM for hierarchical temporal modulation, enabling effective modeling of speech signals with multi-scale dynamics.
\item \textbf{Scalability and efficiency:} We demonstrate that IML-Spikeformer scales to large speech processing tasks, achieving performance comparable to ANN-based Transformers while substantially reducing energy consumption across model sizes, highlighting its promise for energy-efficient speech processing.
\end{itemize}

The rest of the paper is organized as follows. Section \ref{sec: related works} reviews the related works. Section \ref{sec: preliminaries} provides the preliminaries on spiking neuron model and spiking Transformer. Section \ref{sec: IMLS} and Section \ref{sec: archi design} present our core contributions: the IMLS firing mechanism and the HD-RepSSA module, respectively.  Section \ref{sec: exp setup} details our experimental design and settings.  Section \ref{sec: exp results} presents experimental results demonstrating the effectiveness and efficiency of our IML-Spikeformer across ASR, speaker identification, and speaker verification tasks. Finally, Section \ref{sec: conclusion} summarizes this work.

\section{Related works}
\label{sec: related works}
We start by reviewing the existing SNN models for speech processing, which set the stage for the proposed IML-Spikeformer architecture.

\subsection{Spiking Neural Networks for Speech Processing}
SNNs offer the promises of energy efficiency, rapid response, and inherent robustness to noise perturbations. Motivated by these, various SNN architectures have been proposed, such as Spiking Multilayer Perceptrons (SMLP)\cite{bittar2022surrogate, song2024ed}, Spiking Recurrent Neural Networks (SRNN)\cite{bellec2018long, yin2021accurate}, and Spiking Convolutional Neural Networks (SCNN)\cite{yilmaz2020deep, yang2022deep}. These SNN architectures have demonstrated performance comparable to that of conventional ANNs across various speech processing tasks, including KWS\cite{wu2018biologically, Zhang_Yang_Ma_Wu_Li_Tan_2024}, acoustic event and sound classification~\cite{wu2018spiking, pan2020efficient}, ASR\cite{wu2020deep, bittar2022surrogate, ponghiran2022spiking}, sound source localization\cite{li2021axonal,zhang2024spike}, speaker identification~\cite{song2024spiking}, and speech enhancement~\cite{hao2024towards, hao2024audio, du2024spiking}.

Despite these advancements, a significant performance gap persists between SNN models and state-of-the-art ANN models in large-scale speech processing tasks. This performance gap primarily stems from previous SNN architectures' difficulty scaling to larger networks \cite{fang2021deep}, constraining their ability to process complex temporal dependencies and large data scale. Recently, spiking Transformer architecture was a response to the research problem. It shown promising performance on challenging computer vision tasks~\cite{zhou2022spikformer, wang2023complex, yao2024spikev2}. However, our preliminary studies indicate that existing spiking Transformer architectures experience significant performance degradation when applied directly to speech processing tasks where there is a need to process speech signals of variable lengths. This is primarily due to the inability of batch normalization in this condition. We consider IML-Spikeformer as an alternative in this paper.

\subsection{Training Strategy for Large-Scale SNNs}
In general, there are two primary approaches for training large-scale SNNs: Spatio-Temporal Backpropagation (STBP)\cite{yao2023spike, zhou2022spikformer, yao2025scaling} and ANN-to-SNN conversion (ANN2SNN)\cite{wu2021progressive, wang2023complex, bal2024spikingbert}. 

STBP explicitly unrolls the network dynamics of SNNs across \(T\) timesteps and \(L\) layers, applying backpropagation through time (BPTT)\cite{wu2018spatio}. This multi-timestep simulation incurs \(O(L \times T)\) memory complexity to store the neuronal states necessary for gradient computation, as well as \(O(T)\) time complexity due to its inherently iterative processing. On the other hand, ANN2SNN methods convert a pre-trained ANN into SNN via rate-coding approximation, effectively circumventing the training overhead since the ANN2SNN conversion only need finetuning for few epochs. However, such technique necessitates many timesteps (like 16)\cite{wu2021progressive, bal2024spikingbert} to accurately approximate the activations of the ANN with spike firing rate, resulting in high computational costs during inference. Both methods introduce substantial computational overhead—STBP during training and ANN2SNN during inference—making the implementation of large-scale SNNs more expensive.

To resolve these limitations, we propose the IMLS firing mechanism, which effectively simulates multi-timestep neuronal firing within a single training timestep, thereby substantially enhancing the training efficiency of large-scale SNN. Also, our IMLS firing mechanism only need fewer timesteps(like 4 or 6) comparing to the ANN2SNN methods, enabling energy efficient and low-latency inference.

\section{Preliminaries}
\label{sec: preliminaries}
We next introduce the basic concepts of spiking neuron model and spiking Transformer, which are essential for understanding the proposed IML-Spikeformer architecture subsequently.

\subsection{Spiking Neuron Models}
\label{sec: SNN}
Spiking neuron models are computational abstractions motivated by the understanding of biological neurons, acting as the basic computational units of SNNs. Among these, the Leaky Integrate-and-Fire (LIF) model~\cite{maass1997networks} is widely adopted for constructing large-scale SNNs due to its mathematical tractability. The neuronal dynamics of the LIF model can be described by the following discrete-time formulation:

\begin{equation}
v^l[t] = \beta \hspace{1pt} v^l[t - 1] + x^{l-1}[t]- \theta s^l[t-1],
\label{eq: LIF1}
\end{equation}
\begin{equation}
s^l[t] = \mathcal{F}(v^l[t]) =
\begin{cases} 
0, &  v^l[t] < \theta \\ 
1, &  v^l[t] \geq \theta 
\end{cases}
\label{equ: step funciton}
\end{equation}

where $t$ denotes the timestep, $\theta$ is the firing threshold, $x^{l-1}[t]$ is the pre-synaptic input from layer $l-1$, and $v^l[t]$ and $s^l[t]$ denote the membrane potential and output spike in layer $l$, respectively. Equations \ref{eq: LIF1}-\ref{equ: step funciton} describe three fundamental processes of the spiking neuron: leakage \& integration, reset, and firing.

\noindent \textbf{Leakage \& Integration}: This process defines two essential dynamics within a spiking neuron: the decay of information according to a leaky factor $0\leq \beta \leq 1$ and the integration of the pre-synaptic input $x^{l-1}[t]$. When $\beta = 1$, the neuron functions as an Integrate-and-Fire (IF) neuron with no information leakage between timesteps; otherwise, it operates as a LIF neuron.

\noindent \textbf{Reset}: After integration, the membrane potential undergoes soft reset by subtracting $\theta$ from neurons that fired in the previous timestep, as represented by the last term in Eq.~\ref{eq: LIF1}.

\noindent \textbf{Firing}: The function $\mathcal{F}(\cdot)$ represents the spike firing mechanism. When the membrane potential $v^l[t]$ surpasses the firing threshold $\theta$, an output spike $s^l[t] = 1$ is generated; otherwise, $s^l[t] = 0$.

\subsection{Spiking Transformers}
\label{sec: Spiking Transformer}
The spiking Transformer adapts the conventional Transformer architecture to enhance for computational efficiency. The family of spike-driven transformers is an example~\cite{yao2023spike, yao2024spikev2, yao2025scaling}. It typically comprises two key modules: Spike-driven Self-Attention (SDSA) and a Spiking Channel Multi-Layer Perceptron (ChannelMLP)\cite{yao2023spike}. An input sequence $X=\{x[1], x[2], \dots,x[T]\}$ over whole time window is processed by these two modules,
\begin{equation}
\begin{aligned}
    X' &= \text{SDSA}(X) + X, \\X'' &= \text{ChannelMLP}(X') + X'.
\end{aligned}
\label{equ: transformer_updates}
\end{equation}

\begin{figure*}[!t]
\centering
\includegraphics[width=\linewidth]{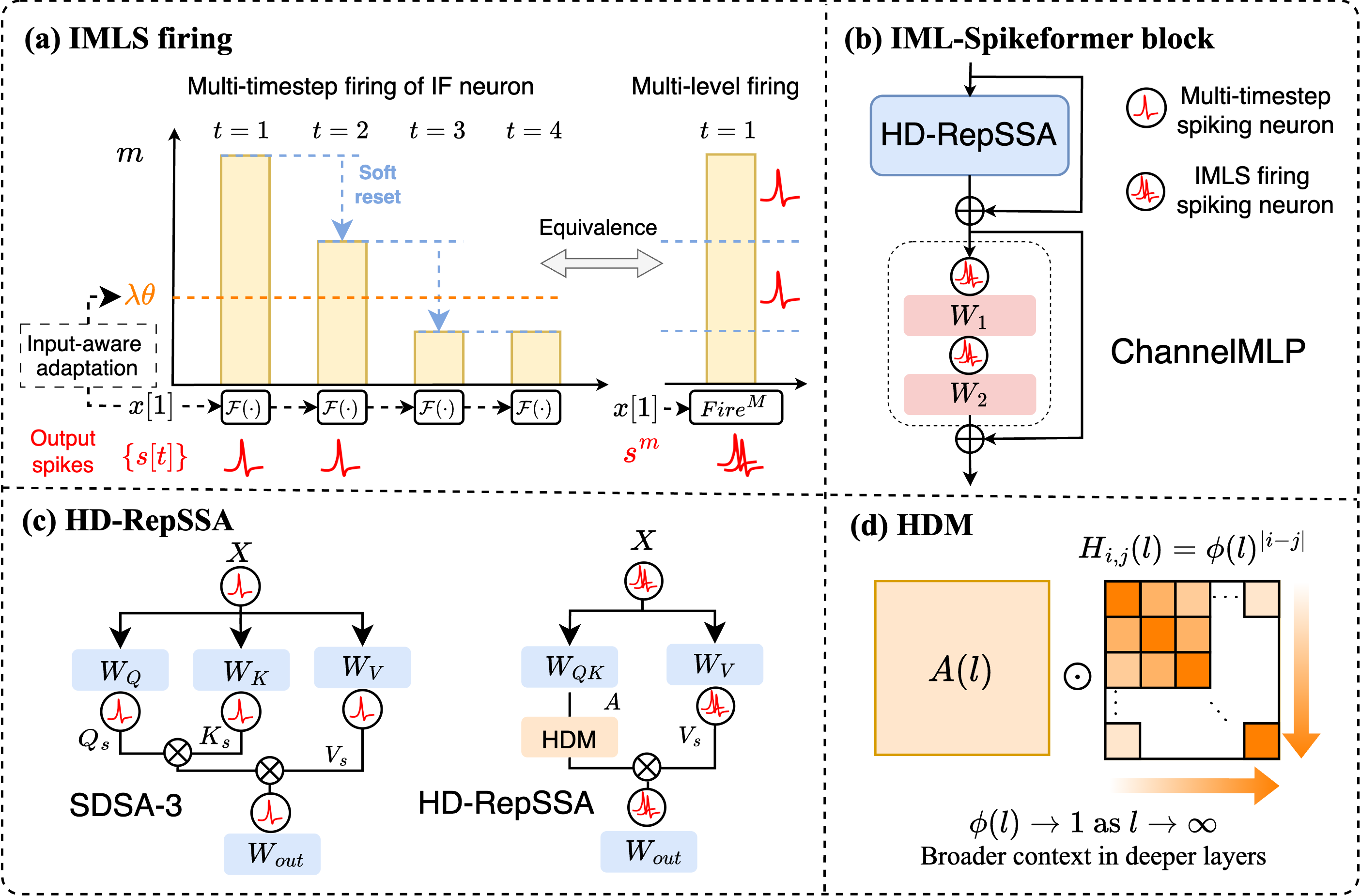}
\caption{Overview of the proposed IML-Spikeformer architecture. (a) When only receive the pre-synaptic input in the first timestep $x[1]$, the membrane potential $m$ iterative update of IF neuron is shown (left). The IMLS firing mechanism offers an equivalence between the output spike train $\{s[t]\}$ generated by the multi-timestep firing of an IF neuron and multi-level spike representation $s^M$ (right) in a single timestep. The threshold $\lambda\theta$ dynamically adjust through the input-aware adaptive scaling factor $\lambda$, which modulates the base threshold $\theta$ in response to the pre-synaptic inputs $x[1]$.
(b) An IML-Spikeformer block consists of two primary components: a HD-RepSSA module and a ChannelMLP module, where the IMLS firing spiking neurons are used for the spiking neuron layers $\mathcal{SN}(\cdot)$ in both modules. The IMLS firing spiking neuron use multi-level spike for efficient training, and converted to equivalent binary spike train for spike-driven inference.
(c) Comparison between the SDSA-3 approach \cite{yao2024spikev2} with multi-timestep spiking neurons (left) and the proposed HD-RepSSA method (right). In the HD-RepSSA module, attention maps are calculated using re-parameterized weight $W_{QK}$ and modulated by the HDM.
(d) The attention map $A(l)$ in layer $l$ is modulated by a HDM $H(l)$ governed by a layer-specific decay factor $\phi(l)$, which progressively increases (approaching 1) in deeper layers. The color intensity in the visualization corresponds to value magnitude, with deeper orange indicating values closer to 1. }
\label{Fig: overview}
\end{figure*}

Specifically, the SDSA module concatenates a self-attention mechanism with a spiking neuron layer, denoted as $\mathcal{SN}(\cdot)$. This spiking neuron layer converts floating-point inputs into binary spikes. Taking SDSA-3~\cite{yao2024spikev2} (shown in Fig.\ref{Fig: overview} (c) left) as an example:
\begin{equation}
\begin{aligned}
    \mathbf{Q} = XW_Q, \quad&\mathbf{K} = XW_K, \quad\mathbf{V} = XW_V,\\
    \text{SDSA}(\mathbf{Q}, \mathbf{K}, \mathbf{V}) &= \mathcal{SN}(\mathbf{Q_s}\mathbf{K_s}^T\mathbf{V_s})W_{\text{out}}, 
\end{aligned}
\end{equation}
where  $W_Q, W_K, W_V$ represent the query, key, value transformation matrices. The query is defined as $\mathbf{Q_s}=\mathcal{SN}(\text{BN}(\mathbf{Q}))$, with $\mathbf{K_s}$ and $\mathbf{V_s}$ defined analogously. The SDSA-3 implements linear attention through the direct multiplication of $\mathbf{Q_s}\mathbf{K_s}^T\mathbf{V_s}$, which can be computed as $\mathbf{Q_s}(\mathbf{K_s}^T\mathbf{V_s})$, achieving linear computational complexity with respect to sequence length and thereby enabling efficient long sequence modeling. $\text{BN}(\cdot)$ denotes the Batch Normalization layer, which is applied to normalize pre-synaptic inputs, facilitating stable information transmission. The results of QKV multiplication is then projected with a linear layer with weight $W_\text{out}$ for the output of SDSA-3 module. Just like the MLP module in ANN transformers, the ChannelMLP module, a two-layer spiking MLP, is applied to spiking Transformer to capture the complex information across channels, formulated as:
\begin{align}
    \text{ChannelMLP}(X) &= \mathcal{SN}(\mathcal{SN}(X)W_1)W_2.
\end{align}
where $W_1, W_2$ are two learnable weights of the spiking MLP.

\section{Input-Aware Multi-Level Spike Firing (IMLS)}
\label{sec: IMLS}
The spiking Transformer architecture that adopts multi-timestep firing mechanism, e.g. STBP, introduces significant computational overhead. Furthermore, such architecture typically employs a fixed firing threshold for spiking neurons, constraining their ability to represent the dynamically varying range of pre-synaptic inputs. To address these issues, we propose an Input-aware Multi-Level spike (IMLS) firing mechanism to enhance both computational efficiency and representation power.

\subsection{Multi-Level Spike Firing}

To mitigate the training overhead of the multi-timestep firing, we introduce the MLS firing mechanism that performs multi-timestep firing in a single timestep. For an IF neuron with soft reset, the equivalence between its $T$-timestep iterative firing and the proposed MLS firing is established in Fig.~\ref{Fig: overview}(a).
Specifically, when receiving a pre-synaptic input only at the first timestep ($X=\{x[1]\}$), the IF neuron continues to emit spikes in subsequent timesteps until its membrane potential $v[t]$ falls below the firing threshold $\theta$ with soft reset, resulting in a spike train $\{s[t]\}, t\in[1, \dots, T]$ (Fig.~\ref{Fig: overview}(a) left show the multi-timestep firing of the IF neuron with $T=4$). 

In contrast, the MLS firing mechanism condenses the multi-timestep iterative process into a single timestep operation. It represents the entire spike train from the IF neuron $\{s[t]\}$ as a single multi-level spike $s^M =\sum^{T}_{t=1}s[t]$ (Fig.~\ref{Fig: overview}(a) right), where the magnitude directly corresponds to the total number of spikes that would have been fired through the complete multi-timestep firing process. The multi-level spike can be formulated as:
\begin{equation}
\begin{aligned}
    s^M &=
    \begin{cases} 
    0, & \quad v[1] < \theta \\[6pt]
    n, & \quad (n-1)\theta \leq v[1] < n\theta \\[6pt]
    T, & \quad (T-1)\theta \leq v[1]
    \end{cases}\\[10pt]
    &= \mathcal{F}^M(v[1], \theta) = \lfloor\text{clip}(\frac{v[1]}{\theta}, 0, T)\rfloor,
\end{aligned}
\end{equation}
where $n \in {1, \dots, T-1}$, $\lfloor \cdot \rfloor$ denotes the floor function, and $v[1]$ represents the membrane potential resulting from the integration of pre-synaptic input $ x[1]$ at the initial timestep. 

This multi-level spike representation in MLS shares similarities with burst coding~\cite{izhikevich2003bursts}. Recent implementations~\cite{park2019fast, li2022efficient, chen2023hybrid} also employ spike magnitude to encode enhanced neural activity beyond binary spikes. However, the fundamental difference lies in their temporal dynamics: while burst coding only considers the number of spikes within a time interval, MLS establishes a mathematical equivalence between multi-timestep IF neuron firing and multi-level spike values. This equivalence (shown in Fig. 1(a)) enables seamless conversion from integer spike computing to binary spike-driven computing during inference.

Since the $\lfloor \cdot \rfloor$ is non-differentiable at the boundaries, we leverage the Generalized Straight-Through Estimator (G-STE)~\cite{liu2022nonuniform} to approximate the gradient, yielding:
\begin{equation}
\label{eq: MLS back}
    \begin{aligned}
 \frac{\partial s^M}{\partial v[1]}& = \mathbb{E}_{s^{M}} \left[ \frac{\partial s^M}{\partial v[1]} \right] = \frac{\partial}{\partial v[1]} \mathbb{E}[s^M]\\ &= \frac{\partial \mathrm{clip}(\frac{v[1]}{\theta}, 0, T )}{\partial v[1]}
=\begin{cases}
\frac{1}{\theta}, & \text{if } 0 \leq v[1] \leq \theta T \\
0. & \text{Otherwise}
\end{cases}
\end{aligned}
\end{equation}

The proposed MLS establishes a memory-time tradeoff in training and inference: during GPU training, MLS utilizes multi-level spike $s^M$ that compresses a spike train into a single timestep, enabling efficient training. During neuromorphic inference, MLS converts the multi-level spikes $s^M$ into equivalent binary spike trains $\{s[t]\}$, preserving the spike-driven computation. This transformation maintains mathematical equivalence as follows,
\begin{equation}
\begin{aligned}
s^M W = \lbrack \sum_{t=1}^{T}s[t]\rbrack W= \sum_{t=1}^{T} s[t]W,
\end{aligned}
\end{equation}
where dense matrix operations $s^M W$ during training become sparse accumulation $s[t]W$ during neuromorphic deployment. By strategically allocating computational resources, prioritizing temporal compression during training and spike-driven computations during inference, the MLS framework allows for memory and time saving during training while enabling efficient neuromorphic hardware deployment.

The MLS firing mechanism is also compatible with neuromorphic chips. In practice, an MLS neuron generating a multi-level spike of value $k$ is implemented as emitting $k$ consecutive binary spikes within a short time window, like burst coding\cite{izhikevich2003bursts} and graded spike firing\cite{shrestha2024efficient}. Current neuromorphic chips like Intel's Loihi \cite{davies2018loihi} and Speck \cite{yao2024spike} support this implementation strategy, making the MLS architecture naturally suited for energy-efficient neuromorphic deployment.

\subsection{Input-aware Multi-Level Spike Firing}
In the previous studies~\cite{zhou2022spikformer, yao2023spike, yao2024spikev2}, the Batch Normalization(BN) layers are employed in spiking Transformer for vision tasks to normalize pre-synaptic inputs, ensuring the membrane potential remains within an appropriate range to sustain stable firing rates. However, BN becomes problematic for variable-length speech sequences because it computes statistics across the batch dimension, leading to inconsistent normalization when sequence lengths vary significantly~\cite{wang2022understanding}. This statistical inconsistency destabilizes training convergence. Conversely, removing BN entirely may result in unbounded pre-synaptic inputs that cause erratic firing patterns, especially in deeper SNNs.

For stable spike firing, we propose Input-aware Multi-Level Spike (IMLS) firing mechanism, which extends the MLS firing mechanism with adaptive thresholds that automatically adjust to pre-synaptic input, allowing our model to maintain stable spike firing patterns even without explicit normalization like BN. The IMLS firing mechanism incorporates a channel-wise adaptive scaling factor $\lambda \in \mathbb{R}^C$ for the firing threshold $\theta$. Unlike fixed thresholds, the adaptive threshold $\lambda\theta$ dynamically adjusts in response to the statistical distribution of pre-synaptic inputs $x[1]$.

During training, for the $i$-th pre-synaptic input batch $x^i[1] \in \mathbb{R}^{B \times L \times C}$, where $B$, $L$, and $C$ denote the batch size, sequence length, and number of channels respectively, $\lambda$ is dynamically updated based on $\Lambda_i$. Here, $\Lambda_i$ represents the maximum pre-synaptic inputs in each channel:
\begin{equation}
\begin{aligned}
\Lambda_i &= \max_{b\in[1,B],l\in[1,L]} x^i_{b,l,:}[1], 
\hspace{3pt} \lambda = \frac{T}{\Lambda_i},
\end{aligned}
\end{equation}
where $T$ represents the time window and the division is performed element-wise.

However, computing the maximum operation for each inference imposes additional computational overhead, especially in neuromorphic devices. To mitigate this, we use a fixed scaling factor $\tilde{\lambda}$ with a running average $\tilde{\Lambda}$ during training with a momentum parameter $\alpha$:
\begin{equation}
    \begin{aligned}
        \tilde{\Lambda} = (1-\alpha) \cdot \tilde{\Lambda} + \alpha \cdot \Lambda_i, 
        \hspace{3pt} \tilde{\lambda} &= \frac{T}{\tilde{\Lambda}}.
    \end{aligned}
\end{equation}
During inference, this scaling factor $\tilde{\Lambda}$ only updated in training and fixed in inference is utilized to modulate the threshold, thereby eliminating the computational overhead while preserving the adaptive properties.

In summary, the input-aware threshold adaptation mechanism extends the dynamic range of spiking neurons by dynamically adjusting thresholds relative to input intensities, enabling informative spike responses across diverse signal distribution. This adaptability stabilizes firing rates despite varying input distributions, preventing both neuronal saturation and silence that commonly occur with fixed thresholds, serving as an implicit normalization mechanism. Our IMLS firing mechanism simultaneously achieves enhanced representational capacity through stable, normalized neural firing and improved computational efficiency via single-timestep processing, establishing IMLS as a robust foundation for training spiking Transformer architectures in large-scale speech processing tasks.

\section{Hierarchical Decay Re-parametrized Spiking Self-Attention}
\label{sec: archi design}
As illustrated in Fig. \ref{Fig: overview} (b), each IML-Spikeformer block is composed of two key components: the HD-RepSSA module for token mixing and the Spiking ChannelMLP for channel mixing. At the core of IML-Spikeformer is the HD-RepSSA (shown in Fig. \ref{Fig: overview} (c)) — a novel spiking self-attention mechanism that precisely captures the hierarchical temporal dependencies inherent in speech signals while preserving the energy-efficient, spike-driven computation paradigm. The spiking neuron layers in HD-RepSSA $\mathcal{SN}(\cdot)$ utilize the IMLS firing spiking neurons introduced in the previous section, which use multi-level spike in training and binary spike train for spike driven inference.

\subsection{Re-parametrized Spiking Self-Attention}
In the vanilla self-attention mechanism from Transformer~\cite{vaswani2017attention}, the attention map is computed as the matrix product of the query and key: $\mathbf{A} = \mathbf{Q}\mathbf{K}^\mathrm{T}$. This operation effectively measures token similarity, particularly when $\mathbf{Q}$ and $\mathbf{K}$ are continuous-valued. However, in spiking self-attention modules like SDSA-3, $\mathbf{Q}$ and $\mathbf{K}$ are converted to spike matrices $\mathbf{Q}_s$ and $\mathbf{K}_s$ via $\mathcal{SN}(\cdot)$. This conversion brings a significant limitation: the resulting attention map $\mathbf{A}_s = \mathbf{Q}_s\mathbf{K}_s^T$ struggles to accurately capture temporal relationships between tokens precisely.

\begin{figure}[!t]
\centering
\includegraphics[width=\linewidth]{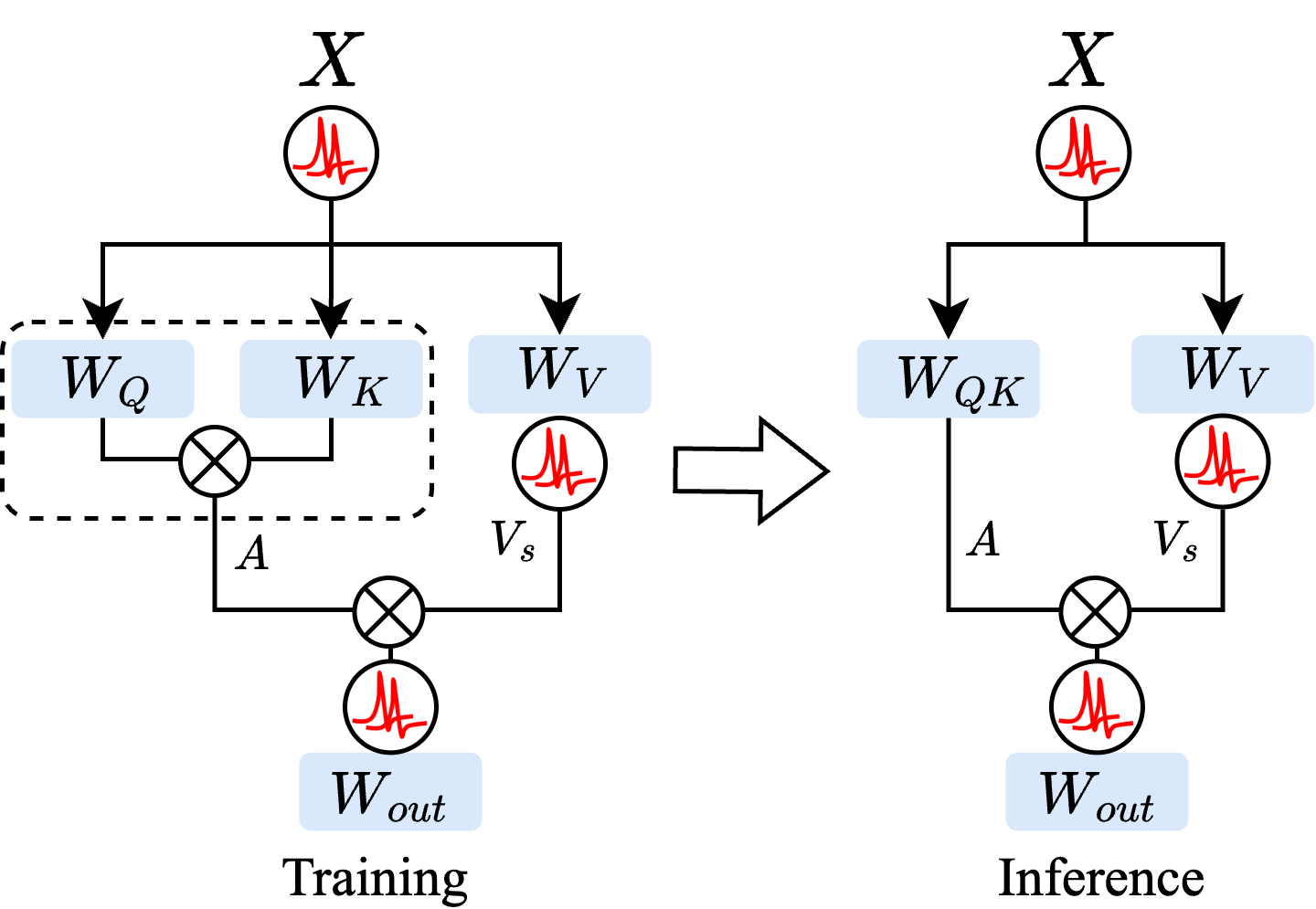}
\caption{Diagram illustrating the RepSSA module. During inference, the weight matrices $W_Q$ and $W_K$ are re-parametrized into a single matrix $W_{QK}$ to maintain spike-driven computation.} 
\label{fig1}
\end{figure}

To address this limitation, we introduce a novel RepSSA mechanism, which utilizes continuous-valued matrices for attention map calculation during training, while employing a re-parametrization technique to ensure spike-driven computations are preserved during inference:
\begin{equation}
    \begin{aligned}
        \mathbf{A} &=
        \begin{cases}
            X W_Q W_K^T X^T, &\text{Training} \\
            X W_{QK} X^T, &\text{Inference}
        \end{cases} \\[6pt]
        \mathbf{V_s} &= \mathcal{SN}(XW_V),
    \end{aligned}
    \label{eq: Rep attention}
\end{equation}
where $X$ denotes the binary spike input. In the inference stage, we fuse the query and key transformation weights $W_Q, W_K$ as $W_{QK} = W_Q W_K^T$. Grounded on this re-parametrization, we define two versions of RepSSA as:
\begin{equation}
\begin{aligned}
\text{RepSSA}_L(\mathbf{A}, \mathbf{V_s}) &= \mathcal{SN}(\mathbf{A} \mathbf{V_s}) W_{out},\\
\text{RepSSA}_S(\mathbf{A}, \mathbf{V_s}) &= \mathcal{SN}(\text{Softmax}(\frac{\mathbf{A}}{\sqrt{d_k}})\mathbf{V_s})W_{out},
\end{aligned}
\end{equation}
where RepSSA$_L$ and RepSSA$_S$ respectively denote linear and softmax spiking self-attentions. $d_k$ denotes the dimension of $\mathbf{K}$. 

Since the token sequence length $N$ in our experiments remains consistently smaller than the feature dimension $D$, linear attention, with a linear computational complexity of $O(ND^2)$ of sequence length $N$, offers no theoretical advantage over softmax attention, which has a complexity of $O(N^2D)$. Moreover, the linear attention mechanism introduces notable challenges in large-scale spiking Transformers, including performance degradation — phenomena primarily attributed to unbounded gradient issues~\cite{qin2022devil}. Given these theoretical and practical considerations, we primarily adopt RepSSA$_S$ in the proposed IML-Spikeformer. Nevertheless, we also evaluate RepSSA$_L$ to explore the potential of linear spiking self-attention for speech processing tasks.

\subsection{Hierarchical Decay Mask}
Speech signals inherently exhibit multi-scale temporal structures, organized hierarchically from phonemes to complete utterances. Neurophysiological studies have demonstrated that auditory cortical processing employs progressively longer temporal integration windows as signals propagate through the cortical hierarchy \cite{hickok2007cortical, rauschecker2009maps}. Drawing inspiration from this biological principle, we design a hierarchical decay mask (HDM) $\mathbf{H}$ to model multi-scale temporal dependencies by modulating the attenuation rate between tokens across different layers (shown in Fig.\ref{Fig: overview} (d)). For two tokens at positions $i$ and $j$ in layer $l$, the corresponding attenuation rate is defined as:
\begin{equation}
    \mathbf{H}_{i,j}(l)=\phi(l)^{|i-j|},
\end{equation}
where $\phi(l)$ represents the layer-specific decay function, formulated as $\phi(l) = 1 - 2^{-5 - l}$. The term $|i-j|$ denotes the absolute positional distance between the tokens. The decay factor increases as the layer depth increases, enforcing a structured attention modulation strategy: shallow layers apply more pronounced attenuation to prioritize local interactions, while deeper layers retain decay values near 1, preserving long-range dependencies crucial for capturing global context.

The HDM can be integrated into the proposed RepSSA through element-wise multiplication with the attention map, creating a biologically-inspired attention mechanism that inherently captures multi-scale temporal dependencies. This integration forms the core component of IML-Spikeformer, which is termed HD-RepSSA, formulated as:
\begin{equation}
\text{HD-RepSSA}(\mathbf{A}, \mathbf{V_s})=\text{RepSSA}(\mathbf{A'}, \mathbf{V_s}), \hspace{3pt} \mathbf{A'}= \mathbf{A} \odot \mathbf{H}.
\label{eq: hdrepssa}
\end{equation}
where $\odot$ is hadamard product.

In general, HDM models the multi-scale temporal dependencies inherent in speech processing, simultaneously capturing fine-grained local acoustic features while representing long-term temporal structures. This dual-scale capability enables IML-Spikeformer to dynamically adapt its processing across multiple temporal resolutions, effectively addressing the hierarchical organization that characterizes natural speech signals. The integration in Eq.~\eqref{eq: hdrepssa} is applied to two RepSSA variants, resulting in HD-RepSSA$_S$ and HD-RepSSA$_L$, respectively.

\section{Experimental setup}
\label{sec: exp setup}
We now provide an overview of the experimental setup. We first outline the configurations for three speech processing tasks: automatic speech recognition (ASR), speaker identification, and speaker verification. We then introduce the baseline models for each task. Finally, we describe how we quantitatively estimate the energy consumption during model inference. 

\subsection{Task Configurations}
To assess the effectiveness and efficiency of the proposed IML-Spikeformer model, we conducted extensive experiments on large-scale speech processing tasks where SNNs historically achieved inferior performance compared to ANNs. In all experiments, IML-Spikeformer is implemented as a direct substitute for ANN Transformer.

\noindent \textbf{ASR:}
The proposed IML-Spikeformer was evaluated on two public ASR databases: AiShell-1~\cite{bu2017aishell}, consisting of 170 hours of Mandarin speech data, and LibriSpeech-960~\cite{panayotov2015librispeech}, comprising approximately 960 hours of English audiobook data. The experimental pipelines follow the established ESPnet ASR recipes~\cite{watanabe2018espnet} for data preparation, model training, and evaluation protocols. We employ ANN Transformers, IML-Spikeformer, and other baseline models (detailed in Section \ref{baseline model}) as encoders, with a fixed 6-layer Transformer decoder without language model across all experiments to ensure fair comparison.

For evaluation, we assess model performance using Character Error Rate (CER) on AiShell-1's development and test sets (labeled as "dev" and "test" in Table \ref{tab:librispeech960}). For LibriSpeech-960, we evaluate using Word Error Rate (WER) on both clean and other subsets of the development and test sets. The clean subset contains high-quality recordings with minimal background noise and clear pronunciation, while the other subset includes more challenging audio with background noise, varied recording conditions, and diverse speaker accents, providing comprehensive assessment across different acoustic conditions.

In data preprocessing, 80-dimensional log Mel filterbank features represent the acoustic inputs. Data augmentation techniques are systematically applied for improved robustness: SpecAugment~\cite{park2019specaugment} for frequency and time masking, and speed perturbation for temporal variability. The input features are subsequently downsampled through two consecutive 3×3 convolutional layers with stride 2, reducing the temporal dimensionality of the original speech inputs.

\noindent \textbf{Speaker Identification and Verification:}
Experiments were also conducted on the VoxCeleb dataset, which includes VoxCeleb1~\cite{nagrani2017voxceleb} and VoxCeleb2~\cite{chung2018voxceleb2}, containing over 2,000 hours of speech from interview videos on YouTube. Specifically, VoxCeleb1 consists of more than 100,000 utterances from 1,251 speakers, while VoxCeleb2 features over 1,000,000 utterances from 6,112 speakers.
For speaker identification, we follow the official partition of VoxCeleb1, where the training set includes 138,361 utterances, and the test set contains 8,251 utterances from the 1,251 speakers. In the speaker verification experiments, the models were trained independently on VoxCeleb1 and VoxCeleb2 development sets, with evaluation performed on the VoxCeleb1-O test partition. 

In all speaker identification and verification experiments, we implemented the pipeline with Transformer-based embedding extractors. All three approaches, namely ANN Transformer, IML-Spikeformer, and spike-driven Transformer baseline, maintain the same architectural configurations and training protocols as outlined in ~\cite{zhang2022mfa}.

\subsection{Baseline Models}
\label{baseline model}
We compare the effectiveness and energy efficiency of IML-Spikeformer against a range of ANN and SNN baseline models for the aforementioned speech tasks. For ANN baselines, we adopt the vanilla Transformer architecture for all three tasks. Additionally, we include an RNN-based model in ASR, that has a VGG network for feature extraction followed by a 3-layer bidirectional LSTM encoder with 1024 hidden and output dimensions, referred to as VGG-BiLSTM.

Due to the limited availability of directly comparable SNN baselines in large-scale speech processing, we re-implement several state-of-the-art SNN models as comparative baselines. We include the Spike-driven Transformer with SDSA-3~\cite{yao2024spikev2}, the latest spiking Transformer architecture, with batch normalization layers removed to ensure convergence. We also implement two 3-layer SMLPs with 1024 hidden and output dimensions as encoders for ASR, incorporating either vanilla LIF neuron models or Gated Spiking Units (GSU)~\cite{hao2024towards}. Notably, GSU enhances sequential modeling through its built-in gating mechanism, demonstrating superior performance in speech enhancement tasks~\cite{timcheck2023intel}.

To provide more comprehensive comparisons on the AiShell-1 ASR task, we include additional sequential processing architectures: binary S4D~\cite{stan2024learning} and Spiking Temporal Convolution Network (Spiking TCN)~\cite{bai2018empirical}. These models represent different paradigms for handling temporal dependencies—S4D through structured state-space modeling and Spiking TCN through dilated convolutions with spiking dynamics. For the AiShell-1 task, we implement 6-layer architectures with 680 and 356 hidden dimensions for S4D and Spiking TCN, respectively. 

For speaker identification, we include Spiking-LEAF~\cite{song2024spiking}, the only existing SRNN-based method applied to this task.  For fair comparison, we used the same architecture in the proposed IML-Spikeformer, the Spike-driven Transformer baseline, and the traditional ANN Transformer. All models have the same number of parameters.

\subsection{Energy Consumption Estimation}

To assess the energy efficiency advantages of SNNs over ANNs, we adopt the energy estimation methods from the Intel N-DNS challenge~\cite{timcheck2023intel}. This operation-based energy evaluation methodology is also commonly used in recent neuromorphic benchmarking studies~\cite{timcheck2023intel, yik2025neurobench}. This approach uses the number of effective operations as a proxy for practical energy consumption, enabling reliable assessment of neuromorphic power advantages without requiring deployment on neuromorphic hardware.

Our evaluation framework separately assesses the energy consumption of ANNs and SNNs due to their different computational paradigms. ANNs rely on dense floating-point matrix computations implemented through Multiply-Accumulate (MAC) operations, while SNNs exploit sparse, event-driven processing using Accumulate (AC) operations for synaptic integration and neuronal dynamics. We calculated the detailed energy consumption based on previous studies on 45nm CMOS technology~\cite{horowitz20141}, where one floating-point MAC operation consumes $E_{\text{MAC}} = 4.6$~pJ while one AC operation consumes $E_{\text{AC}} = 0.9$~pJ.

For conventional ANNs, energy consumption scales linearly with the total floating-point computational load:
\begin{equation}
E_{\text{ANN}} = E_{FLOPs} = E_{\text{MAC}} \sum_{l=1}^{L} \text{FLOPs}^{l}
\end{equation}
where the energy consumption of ANNs comes entirely from Floating Point Operations (FLOPs), $\text{FLOPs}^{l}$ represents the floating-point operations in layer $l$, and $L$ denotes the total network depth.

For SNNs, energy consumption comprises two components reflecting Synaptic Operations (SynOPs) and Neuronal Operations (NeuOPs):
\begin{align}
E_{\text{SNN}} &= E_{\text{SynOPs}} + E_{\text{NeuOPs}} \\
E_{\text{SynOPs}} &= E_{\text{AC}} \sum_{l=1}^{L} \text{SynOPs}^{l} \\
E_{\text{NeuOPs}} &= 10 \times E_{\text{AC}} \sum_{l=1}^{L} \mathcal{N}^l
\end{align}
The synaptic operations are quantified as $\text{SynOPs}^l = T \sum_{i} R^l_i \cdot C^l_i$, where $T$ represents the simulation time window, and $R_i^l$ and $C_i^l$ denote the firing rate and incoming synaptic connections of presynaptic neuron $i$ in layer $l$, respectively. The detailed calculations of attention modules' synaptic operations are presented in Table I of the Supplementary Materials. The neuronal operations are quantified by $\mathcal{N}^l$ active neurons per layer, where each neuron operation consumes approximately 10 times the energy of a synaptic operation on the Loihi architecture~\cite{timcheck2023intel}.
\color{black}

\section{Experimental results}
\label{sec: exp results}
\subsection{Main Results}
\label{sec: exp}
In this section, we report the main results of the proposed IML-Spikeformer alongside baseline models for comparison across three tasks: ASR, speaker identification, and speaker verification. We will make our code publicly available after the review process.

\subsubsection{Automatic Speech Recognition:}
Table~\ref{tab:aishell} presents a comprehensive comparison of our IML-Spikeformer against both ANN and SNN baselines on the AiShell-1 ASR task. Our proposed IML-Spikeformer (30.35M parameters, 6 timesteps) achieves a Character Error Rate (CER) of 5.9\% on the test set, outperforming existing SNN approaches by substantial margins. Notably, it reduces the test CER by 9.8\% and 6.7\% compared to the LIF-based model (15.7\% CER) and the GSU baseline~\cite{hao2024towards} (12.6\% CER), respectively. Furthermore, IML-Spikeformer surpasses the Spike-driven Transformer baseline, achieving a 6.0\% absolute CER reduction (from 11.9\% to 5.9\%).
Compared with the ANN Transformer baseline results reported in the ESPnet toolkit~\cite{watanabe2018espnet}, the proposed IML-Spikeformer achieves comparable CER on the test set while delivering substantial computational efficiency benefits—remarkably, a 4.64× reduction in energy consumption.

\begin{table*}
\centering
\caption{Results on AiShell-1 ASR task showing Character Error Rate (CER) on development and test sets, energy consumption, and energy saving ratio for the proposed IML-Spikeformer compared to ANN and SNN baselines. The energy saving ratio is relative to ANN Transformer (denoted as "×1").}
\label{tab:aishell}

\begin{tabular}{cccccccc}
    \toprule
    Model & SNN & Parameters & Timestep & dev & test & Energy(mJ) & Energy Saving\\
    \midrule
    VGG-BiLSTM*\cite{ravanelli2021speechbrain} & \XSolidBrush & 93.26M & 1 & 9.7 & 10.7 & 33.42 & $\times$ 0.31\\
    Binary S4D*~\cite{stan2024learning} & \XSolidBrush & 21.00M & 1 & 10.3 & 11.9 & 1.73 & $\times$ 6.01\\
    \multirow{3}{*}{Transformer\cite{watanabe2018espnet}} & \XSolidBrush & 30.35M & 1 & 5.6 & 5.9 & 10.39 & $\times$ 1 \\
                                                        & \XSolidBrush & 46.11M & 1 & 5.3 & 5.6 & 15.58 & $\times$ 0.67 \\
                                                        & \XSolidBrush & 61.89M & 1 & 5.1 & 5.6 & 20.78 &  $\times$ 0.5 \\
    
    \midrule
    LIF* & \CheckmarkBold & 22.96M & 1 & 13.8 & 15.7 & 0.73& $\times$ 14.19 \\
    spiking TCN*~\cite{bai2018empirical} & \CheckmarkBold & 23.35M & 1 & 11.4 & 13.7 & 1.62 & $\times$ 6.41\\
    GSU*\cite{hao2024towards} & \CheckmarkBold & 60.74M & 1 & 10.9 & 12.6 & 3.20 & $\times$ 3.25 \\
    Spike-driven Transformer*\cite{yao2023spike} & \CheckmarkBold & 30.35M & 6 & 10.2 & 11.9 & 1.98 & $\times$ 5.24 \\

    \midrule
    \multirow{4}{*}{\textbf{IML-Spikeformer}} & \CheckmarkBold & 30.35M & 4 & 5.8 & 6.2 &1.74  & $\times$ 5.96 \\
                                    & \CheckmarkBold & 30.35M & 6 & 5.5 & 6.0 & 2.24 & $\times$ 4.64 \\
                                    & \CheckmarkBold & 46.11M & 6  & 5.3 & 5.7 &3.12 & $\times$ 3.32 \\
                                    & \CheckmarkBold & 61.89M & 6  & 5.2 & 5.7 &4.03 & $\times$ 2.58 \\
    \midrule
    \multicolumn{8}{l}{* \quad Our reproduced results based on publicly available codebases.} \\
    
\end{tabular}
\end{table*}

For the ASR results on the LibriSpeech-960 dataset, shown in Table~\ref{tab:librispeech960}, IML-Spikeformer achieves a Word Error Rate (WER) of 3.1\% on development set and 3.4\% on the test set — a performance on par with ANN Transformer baselines. Additionally, IML-Spikeformer significantly outperforms SMLPs with the LIF neuron model~\cite{bittar2022surrogate}, alongside the GSU model~\cite{hao2024audio} and spike-driven Transformer baselines, demonstrating its superiority in large-vocabulary ASR tasks.

\begin{table*}
\begin{center}
\caption{Results on LibriSpeech-960 ASR task showing Word Error Rate (WER) on development and test sets, energy consumption, and energy saving ratio for the proposed IML-Spikeformer compared to ANN and SNN baselines. The "dev" and "test" refer to the WER in development and test sets.}
\label{tab:librispeech960}
\begin{tabular}{cccccccccc}
    \toprule
    \multirow{2}{*}{Model} & \multirow{2}{*}{SNN} & \multirow{2}{*}{Parameters} & \multirow{2}{*}{Timestep} & \multicolumn{2}{c}{dev (\%)} & \multicolumn{2}{c}{test (\%)} & \multirow{2}{*}{Energy(mJ)} & \multirow{2}{*}{Energy Saving} \\
    \cmidrule(lr){5-6} \cmidrule(lr){7-8}
    & & & &  clean & other & clean & other & & \\
    \midrule
    VGG-BiLSTM*\cite{ravanelli2021speechbrain} & \XSolidBrush & 202.4M & 1 & 7.2 &18.9 & 7.3 &19.7 & 38.63 & $\times$ 0.85 \\
    Transformer\cite{watanabe2018espnet}  & \XSolidBrush & 99.36M & 1 & \textbf{2.8} & \textbf{7.6} & \textbf{3.2} & \textbf{8.0}& 32.82 & $\times$ 1 \\
    \midrule
    LIF\cite{bittar2022surrogate} & \CheckmarkBold & - & 1 & - & - & 9.94 & - & - & - \\
    GSU*\cite{hao2024towards} & \CheckmarkBold & 185.1M & 1 & 12.4 & 30.4 & 12.8 & 32.1& 8.59  & $\times$  3.82 \\
    \midrule
    \multirow{2}{*}{Spike-driven Transformer*\cite{yao2023spike}} & \CheckmarkBold & 99.36M & 4 & 10.4 &25.3 & 10.5 &25.7 &  4.61 & $\times$ 7.12 \\
                                                                & \CheckmarkBold & 99.36M & 6 & 8.7& 20.7& 8.9 & 22.3&   6.02& $\times$ 5.45 \\
    \midrule
    \multirow{2}{*}{\textbf{IML-Spikeformer}} & \CheckmarkBold & 99.36M & 4  & 3.5& 8.4 & 3.9 & 8.7 &  5.67 & $\times$ 5.79 \\
                                    & \CheckmarkBold & 99.36M & 6 & \underline{3.1}& \underline{8.3}& \underline{3.4}&\textbf{7.9} & 7.60  & $\times$  4.32 \\
    \bottomrule
    \multicolumn{10}{l}{* \quad Our reproduced results based on publicly available codebases. } \\
    \multicolumn{10}{l}{- \quad These results are not publicly available.} \\
    
\end{tabular}
\end{center}
\end{table*}

\subsubsection{Speaker Identification and Verification}
Beyond ASR, IML-Spikeformer is also evaluated on speaker identification and verification tasks, with results illustrated in Table~\ref{tab:Sid} and Table~\ref{tab:ASV}, respectively.
\begin{table}
\begin{center}
\caption{Results on VoxCeleb1 speaker identification task showing test accuracy on the VoxCeleb1 test set and energy consumption for the proposed IML-Spikeformer compared to ANN Transformer and the previous published SNN work\cite{song2024spiking}.}
\label{tab:Sid}
\resizebox{\columnwidth}{!}{
    \begin{tabular}{cccccc}
    \toprule
    Model &Params& SNN& Timestep & Acc.(\%) &Energy(mJ)\\
    \midrule
    Spiking-LEAF\cite{song2024spiking}& 0.9M& \CheckmarkBold & 1 & 30.45& 0.072\\
    \midrule
    \multirow{3}{*}{Transformer*}& 0.6M & \XSolidBrush & 1 &  67.49& 0.65 \\
                                & 1.18M & \XSolidBrush & 1 & 71.31& 1.03  \\
                                & 1.76M & \XSolidBrush & 1 &  73.10& 1.37\\
    \midrule
    \multirow{3}{*}{\textbf{IML-Spikeformer}}& 0.6M& \CheckmarkBold & 6 &  67.43& 0.18 \\
                                & 1.18M& \CheckmarkBold & 6 & 71.83 & 0.24\\
                                & 1.76M& \CheckmarkBold & 6 &  74.34 & 0.34\\
    \midrule
    \multicolumn{6}{l}{* \quad Our reproduced results based on publicly available codebases.} \\
    \end{tabular}
}
\end{center}
\end{table}

As demonstrated in Table~\ref{tab:Sid}, the proposed IML-Spikeformer exhibits robust performance across multiple parameter configurations while maintaining substantial computational efficiency. It is observed that IML-Spikeformer achieves classification accuracies of 67.43\%, 71.83\%, and 74.34\% for model sizes of 0.6M, 1.18M, and 1.76M parameters, respectively. Notably, the largest configuration surpasses its ANN Transformer counterpart, obtaining an improvement of 1.24\% accuracy. In comparison with the SNN baseline, IML-Spikeformer significantly surpasses the SRNN-based implementation presented in~\cite{song2024spiking} (67.43\% versus 30.45\%), despite utilizing fewer parameters. 

\begin{table}
\begin{center}
\caption{Results on speaker verification task showing test accuracy on VoxCeleb1-O test set, energy consumption for the proposed IML-Spikeformer, Transformer, and Spike-driven Transformer baseline\cite{yao2024spikev2}. Results for models trained on VoxCeleb1 and VoxCeleb2 development set are presented.}
\label{tab:ASV}
\resizebox{\columnwidth}{!}{
    \begin{tabular}{ccccc}
    \toprule
    Model & SNN  & Timestep & EER(\%) & Energy(mJ) \\
    \midrule
    \multicolumn{5}{c}{\textbf{Training on VoxCeleb1 dev set}} \\
    \midrule
    Transformer\cite{zhang2022mfa}* & \XSolidBrush  & 1 & 4.44 & 6.85 \\
    Spike-driven Transformer*\cite{yao2024spikev2} & \CheckmarkBold  & 8 & 5.65&  1.55\\
    \multirow{2}{*}{\textbf{IML-Spikeformer}} & \CheckmarkBold & 6 &  4.75 & 1.58 \\
                                & \CheckmarkBold & 8 &  4.47 & 2.37 \\
    \midrule
    \multicolumn{5}{c}{\textbf{Training on VoxCeleb2 dev set}} \\
    \midrule
    Transformer\cite{zhang2022mfa}* & \XSolidBrush  & 1 & 2.66 &  6.85\\
    Spike-driven Transformer*\cite{yao2024spikev2} & \CheckmarkBold  & 8 & 4.53 &  1.62\\
    \multirow{2}{*}{\textbf{IML-Spikeformer}} & \CheckmarkBold & 6 &  3.20 & 1.62\\
                                & \CheckmarkBold & 8 &  2.70 & 2.53\\
    \midrule
    \multicolumn{5}{l}{* \quad Our reproduced results based on publicly available codebases.} \\
    \end{tabular}
}
\end{center}
\end{table}

Table~\ref{tab:ASV} presents speaker verification performance on the VoxCeleb1-O evaluation set. For models trained on the VoxCeleb1 development set, our IML-Spikeformer with 8 timesteps achieves an Equal Error Rate (EER) of 4.47\%, substantially outperforming the Spike-driven Transformer baseline (5.65\%) while approaching the performance of the ANN Transformer (4.44\%). When trained on the larger VoxCeleb2 development set, IML-Spikeformer achieves a competitive 2.70\% EER compared to the ANN Transformer's 2.66\% and significantly surpassing the Spike-driven Transformer baseline (4.53\%). Notably, across all experimental configurations, IML-Spikeformer maintains comparable EER to ANN counterparts while consuming only 36.9\% of the energy (2.53 versus 6.85mJ).

\subsubsection{Scalability of IML-Spikeformer}
As demonstrated in Table \ref{tab:aishell}, we compared our IML-Spikeformer against ANN transformers across three model sizes (30.35M, 46.11M, and 61.89M parameters) on the AISHELL-1 ASR task. The results show that CER decreases consistently for both model types as parameter count increases, with IML-Spikeformer maintaining competitive performance at each scale. Similarly, Table \ref{tab:Sid} reveals that our model achieves comparable or superior performance compared to ANN transformers across all model scales for speaker identification. These consistent results across different tasks and model sizes demonstrate both the parameter scalability and effectiveness of our IML-Spikeformer architecture.

\subsection{Ablation Studies}
As shown in Table \ref{Tab: abla}, we conduct ablation studies on the Aishell-1 ASR task to evaluate the contribution of key components in IML-Spikeformer across three critical dimensions:

\noindent\textbf{Firing methods}: Replacing IMLS firing mechanism with MLS results in substantial performance degradation, increasing CER by 2.4\% on test sets. The performance decline is even more pronounced when switching to iterative multi-timestep firing, with CER increases of 3.2\%. Notably, the negligible energy differences between IMLS, MLS, and multi-timestep firing indicate that IMLS maintains comparable spike firing rate while delivering superior performance through its input-aware threshold adaptation. These results demonstrate the significant performance advantages conferred by the IMLS firing mechanism.

\noindent\textbf{HD-RepSSA components}: We ablate the contributions of individual components within our HD-RepSSA, including RepSSA and HDM. Removing the HDM (HD-RepSSA$_S$ $\rightarrow$ RepSSA$_S$) leads to a modest CER increase of 0.6\% on test set. More significantly, replacing HD-RepSSA$_S$ with standard SDSA increases CER by 1.5\%. These results confirm that while both components contribute positively, the RepSSA module with its enhanced attention map capacity delivers more substantial performance benefits.

\noindent\textbf{Linear attention}: We evaluated linear attention variants to analyze the accuracy-efficiency tradeoff. Switching from HD-RepSSA$_S$ to its linear counterpart (HD-RepSSA$_L$) increases CER by 0.8\% while providing slight energy reduction, demonstrating the classic tradeoff between computational efficiency and model accuracy. The substantial performance gap between HD-RepSSA$_L$ and RepSSA$_L$ further demonstrates the critical role of HDM in maintaining speech processing performance, particularly under linear attention settings where performance degradation is more pronounced.

\begin{table}
\begin{center}
\caption{Ablation studies on IML-Spikeformer for the AiShell-1 ASR task. Performance for various model configurations are presented with relative changes shown in brackets.}
\label{Tab: abla}
\resizebox{\linewidth}{!}{
    \begin{tabular}{lccc}
    \toprule
   Methods &Energy(mJ) & dev & test \\
    \midrule
     IML-Spikeformer & 2.24 & 5.5 & 6.0 \\
    \midrule
    \textbf{Firing methods}&&\\
    IMLS $\rightarrow$ MLS& 2.20(-0.04)&7.6(+2.1) & 8.3(+2.4)\\
    IMLS $\rightarrow$ multi-timestep& 2.16(-0.08)&8.2(+2.7) & 9.1(+3.2) \\
    \midrule
    \textbf{HD-RepSSA components}&&\\
    HD-RepSSA$_S$ $\rightarrow$ RepSSA$_S$&2.19(-0.5)&5.9(+0.4) &6.5(+0.6) \\
    HD-RepSSA$_S$ $\rightarrow$ SDSA-3&1.99(-0.25)&8.1(+2.6) & 9.0(+3.1) \\
    \midrule
    \textbf{Linear attention}&&\\
    HD-RepSSA$_S$ $\rightarrow$ HD-RepSSA$_L$&2.06(-0.18)&6.1(+0.6) &6.7(+0.8) \\
    HD-RepSSA$_S$ $\rightarrow$ RepSSA$_L$&2.01(-0.23)&7.9(+2.4) &8.7(+2.8) \\
    \bottomrule
    \end{tabular}
    }
\end{center}
\end{table}

\subsection{Spike Firing Rate Analysis}
\begin{figure*}[!t]
\centering
\includegraphics[width=\linewidth]{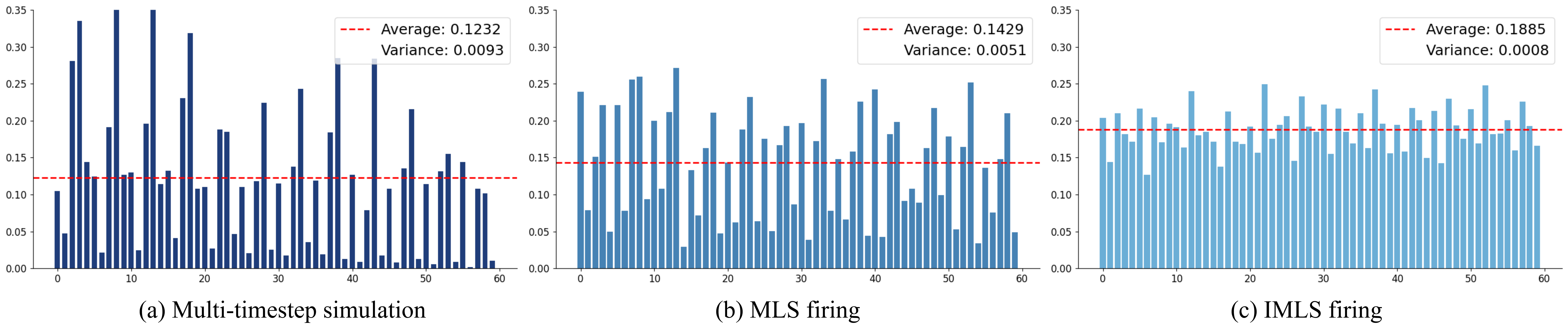}
\caption{Distribution of neuronal spike firing rates across network for three spike firing mechanisms on IML-Spikeformer. X-axis indicates the neuron indices. (a) Vanilla multi-timestep firing exhibits significant proportions of inactive neurons (near-zero firing rates), particularly in deeper layers. (b) IML-Spikeformer with MLS firing mechanism and fixed threshold demonstrates improved but still inconsistent activation patterns across network depth. (c) IML-Spikeformer with IMLS firing mechanism and input-aware threshold adaptation maintains stable firing rate distributions throughout all layers, with substantially reduced neuronal inactivity and more balanced activation patterns.}
\label{Fig: MultiSpike_FR}
\end{figure*}

As evidenced in Table~\ref{Tab: abla}, both MLS and IMLS yield performance improvements over multi-timestep firing, with IMLS demonstrating markedly superior results. To elucidate the underlying mechanism responsible for these performance improvements, we analyze neuronal spike firing rate distributions of each linear layer across network depths for three firing mechanisms on our IML-Spikeformer, as visualized in Fig.~\ref{Fig: MultiSpike_FR}.

Fig.~\ref{Fig: MultiSpike_FR}(a) reveals a critical limitation in the multi-timestep firing: a substantial proportion of spiking neurons exhibit near-zero firing rates across network layers, resulting in sparse activation patterns. This widespread neuronal inactivity substantially constrains the network's representational capacity, as established in previous study \cite{guo2022loss}, directly accounting for the model's performance degradation. Fig.~\ref{Fig: MultiSpike_FR}(b) demonstrates that while MLS firing mechanism with the fixed threshold improves overall performance, it still exhibits considerable proportions of inactive neurons, indicating persistent representational constraints.

In contrast, the proposed IMLS firing mechanism directly addresses these limitations through input-aware threshold adaptation. By continuously monitoring pre-synaptic input distributions and dynamically calibrating neuronal firing thresholds to match these statistics, IMLS maintains appropriate firing rates throughout the network as shown in  Fig.~\ref{Fig: MultiSpike_FR}(c). This adaptive threshold mechanism effectively functions as an implicit normalization operation that eliminates the need for problematic BN layers. The resulting firing rate stabilization ensures consistent information propagation through all network depths, facilitating effective representation transformation across diverse speech patterns and enabling the robust performance scaling observed in our IML-Spikeformer architecture.

\subsection{Attention Maps Analysis}
\begin{figure*}[!t]
\centering
\includegraphics[width=\linewidth]{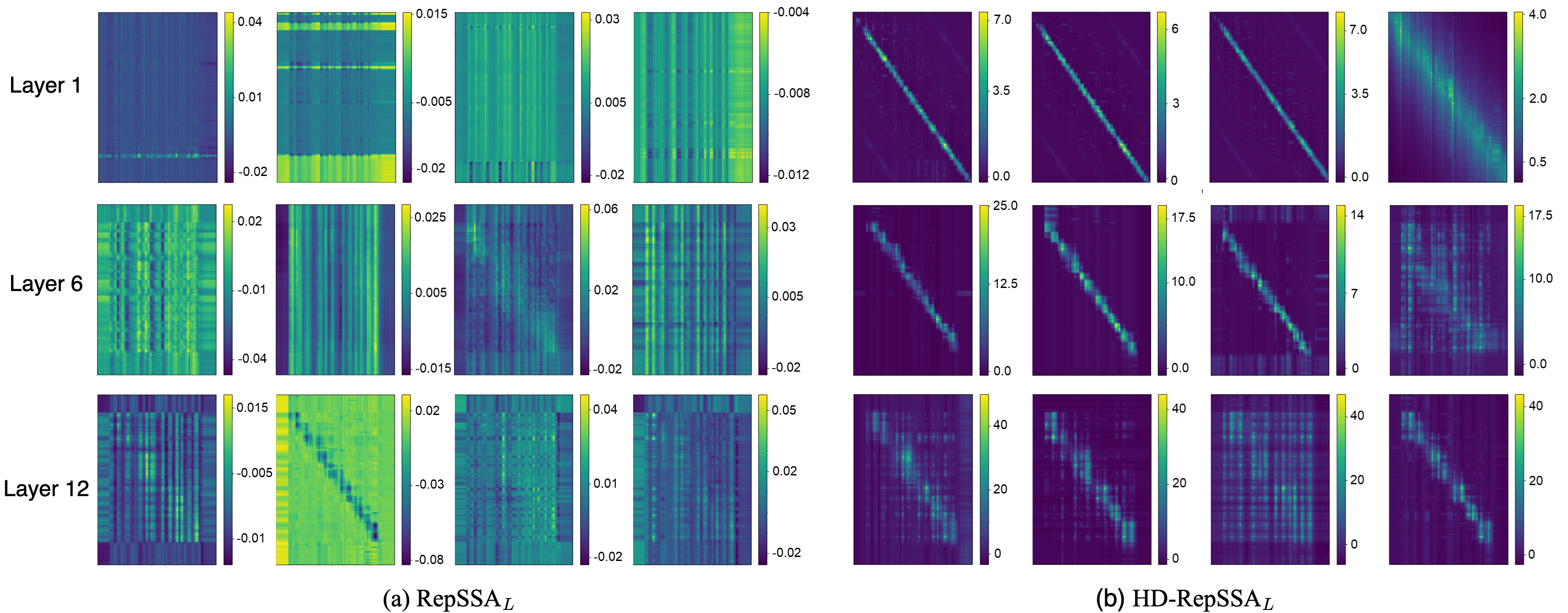}
\caption{Attention map visualization of across network depths in IML-Spikeformer with HD-RepSSA$_L$ and RepSSA$_L$. Attention maps of 4 heads shown at shallow (layer 1), intermediate (layer 6), and deep (layer 12) stages demonstrate how HDM enables hierarchical temporal dependencies, transitioning from local focus in shallow layers to global patterns in deeper layers.}
\label{Fig: vis of attention map}
\end{figure*}
The empirical results presented in Table V demonstrate a notable discrepancy in the performance impact of HDM across softmax and linear attention. When integrated with linear attention mechanisms, HDM yields a substantial CER reduction of 2.3\% (RepSSA$_L$ $\rightarrow$ HD-RepSSA$_L$), whereas the corresponding improvement in softmax-based attention remains comparatively modest at 0.6\% CER reduction (RepSSA$_S$ $\rightarrow$ HD-RepSSA$_S$). This differential impact suggests that HDM provides essential inductive bias that specifically addresses representational limitations inherent in linear attention formulations.

To elucidate the underlying mechanisms, Figure 4 presents a comparative visualization of attention maps across network depth. Figure 4(a) reveals that RepSSA$_L$ without HDM generates attention maps characterized by near-uniform weight distributions across all token pairs. This uniformity persists throughout the network hierarchy (layers 1, 6, and 12), indicating the model's inability to establish differentiated token relationships or develop context-sensitive representational focus, thereby limiting the model's capacity for contextual feature extraction.

In contrast, the HD-RepSSA$_L$ visualization in Figure 4(b) exhibits structured attention maps with layer-dependent characteristics. Early layers (e.g., layer 1) demonstrate concentrated diagonal activation, establishing localized temporal dependencies that correspond to phoneme-level processing. Intermediate and deeper layers (layers 6 and 12) gradually broaden their receptive fields, capturing hierarchically organized speech structures and acoustic features from phonemes to words to sentences. The expanding attention map range directly reflects the increasing receptive field size, where narrow attention spans in early layers capture phoneme-level acoustic details, medium-range attention in intermediate layers integrates phonemic sequences into word representations, and broad attention patterns in deeper layers enable sentence-level contextual understanding. This progressive expansion of the receptive field inherently supports the multi-scale temporal processing crucial for speech tasks, enabling the model to preserve local precision while capturing global context—both essential for effective speech representation. The hierarchical attention structure provides empirical explanation for the performance improvement observed in Table \ref{Tab: abla}, demonstrating how HDM utilizes structured inductive bias that compensates for the representational limitations of linear attention.

\subsection{Training Efficiency Analysis}
\begin{figure}[!t]
\centering
\includegraphics[width=\linewidth]{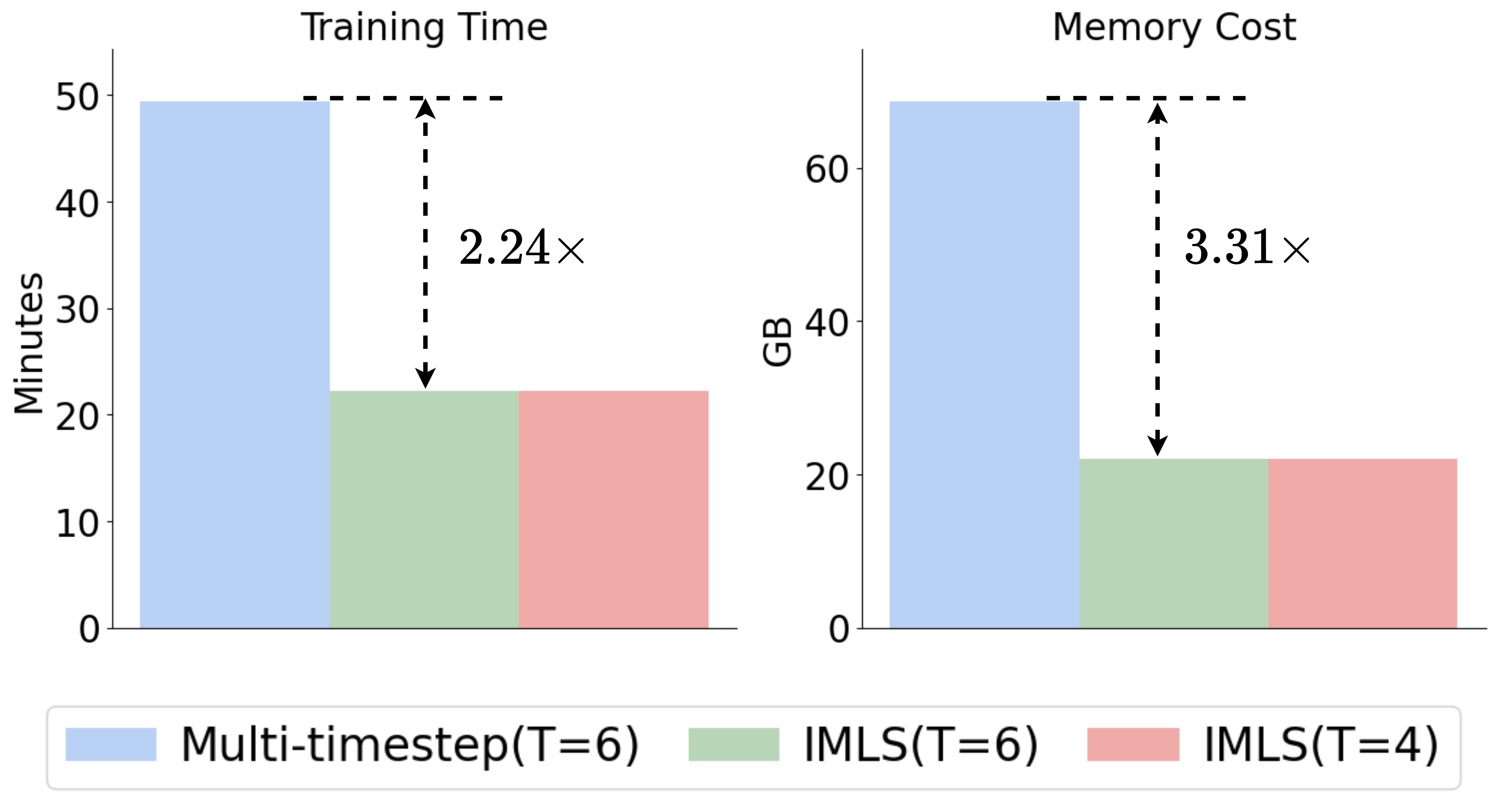}
\caption{The training time per epoch and the GPU memory cost of IML-Spikeformer with the iterative multi-timestep firing and our IMLS firing mechanisms, evaluated on 4 NVIDIA GeForce RTX 3090 Ti GPUs. The IMLS firing mechanism reduces computational overhead with single-timestep training. }
\label{Fig: train cost}
\end{figure}

The IMLS firing mechanism not only stabilizes spike firing but also significantly enhances overall training efficiency and maintained fast convergence speed. Conventional multi-timestep iterative firing requires iterative firing across $T$ timesteps, demanding storage of all intermediate states for BPTT, resulting in memory complexity of $O(L \times T)$ for an $L$-layer network. In contrast, IMLS computes multi-level spikes directly from the membrane potential at the initial timestep, eliminating extended timesteps and intermediate state storage, thereby reducing training and memory complexity to $O(L)$. 

As demonstrated in Fig. \ref{Fig: train cost}, when training IML-Spikeformer on 4 NVIDIA GeForce RTX 3090 Ti GPUs, IMLS achieves a 2.24$\times$ reduction in per-epoch training time and cuts memory costs by 3.31$\times$ compared to iterative multi-timestep firing. Notably, IMLS maintains constant training computational and memory costs regardless of the number of timesteps (T), as evidenced by comparing IMLS with T=4 and T=6. This characteristic makes IMLS especially valuable for our IML-Spikeformer architecture when addressing complex speech processing tasks that benefit from extended $T$ to achieve optimal performance.

To establish the complete training efficiency picture, Fig. 1 in the Supplementary Materials shows that IMLS achieves comparable convergence speed to the multi-timestep baseline, ensuring that the computational optimizations do not come at the cost of learning effectiveness. Consequently, the combination of faster per-epoch training with equivalent convergence speed translates to significantly reduced overall training time to convergence, establishing IMLS as a practically superior solution for efficient SNN training.

\section{Conclusion}
\label{sec: conclusion}
In this paper, we introduced IML-Spikeformer, a novel spiking Transformer architecture designed for large-scale speech processing. Our empirical evaluation confirms that IMLS enhances training efficiency while significantly improving performance and stability through adaptive threshold modulation. Through our proposed HD-RepSSA spiking self-attention module, IML-Spikeformer effectively overcomes the limited representational capacity of conventional spiking self-attention while successfully capturing the hierarchical temporal dependencies characteristic of speech signals. These innovations collectively establish IML-Spikeformer as a promising efficient framework for large-scale speech processing that achieves comparable performance to ANN transformers across ASR, speaker identification, and verification tasks while maintaining the efficiency benefits of SNNs.
\bibliographystyle{IEEEtran}
\bibliography{ref}

\newpage
\onecolumn

\setcounter{page}{1}

\Large
\begin{center}
    {\bf \LARGE Supplementary Materials}\\
    \vspace{0.2cm}
    {\large IML-Spikeformer: Input-aware Multi-Level Spiking Transformer for Speech Processing}\\
	\vspace{0.2cm}
\end{center}
\rule[-0.5pt]{18.1cm}{0.06em}

\parskip=2pt

\small

\parskip=2pt

\subsection{Computing Infrastructure}
All experiments are conducted on Ubuntu 20.04.5 LTS server equipped with NVIDIA GeForce RTX 3090 GPUs (24G Memory), Intel(R) Xeon(R) Platinum 8370C CPU @ 2.80GHz, Pytorch 1.13.0, and CUDA 11.8.

\subsection{Detailed energy consumption}
In this section, we present the detailed energy calculation of attention modules. Table 1 show the $E_{FLOPs}$ of the Vallina Self-Attention(VSA) and $E_{SynOPs}$ of SDSA-3, HD-RepSSA$_S$ and HD-RepSSA$_L$, respectively.

\begin{table}[htbp]
\centering
\caption{Energy consumption of self-attention modules. $T$, $N$, $D$ are simulation timestep, token number and input dimension. $R_C, \hat{R}, R_A$ denote the average spike firing rates of various spike matrices.}
\label{tab:flops_attention}
\resizebox{\linewidth}{!}{
\begin{tabular}{@{}lcccc@{}}
\toprule
 & VSA & SDSA-3 & HD-RepSSA$_S$ & HD-RepSSA$_S$ \\
\midrule
$Q, K, V$ & $3ND^2\cdot E_{MAC}$ & $T \cdot R_C \cdot 3 ND^2\cdot E_{AC}$ & $T \cdot R_C \cdot ND^2\cdot E_{AC}$ & $T \cdot R_C \cdot ND^2\cdot E_{AC}$  \\
$f(Q, K, V)$ & $2N^2D\cdot E_{MAC}$ & $T \cdot \hat{R} \cdot ND^2\cdot E_{AC}$ & $T \cdot \hat{R} \cdot 2N^2D\cdot E_{AC}$ & $T \cdot \hat{R} \cdot 2ND^2\cdot E_{AC}$ \\
HDM & - & - & $T \cdot R_A \cdot N^2\cdot E_{MAC}$ & $T \cdot R_A \cdot N^2\cdot E_{MAC}$ \\
Scale & $N^2\cdot E_{MAC}$ & - & - & - \\
Softmax & $2N^2\cdot E_{MAC}$ & - & $2N^2\cdot E_{MAC}$ & - \\
Linear & $OP_{\text{MLP}}\cdot E_{MAC}$ & $T \cdot R_C \cdot OP_{\text{MLP}}\cdot E_{AC}$ & $T \cdot R_C \cdot OP_{\text{MLP}}\cdot E_{AC}$ & $T \cdot R_C \cdot OP_{\text{MLP}}\cdot E_{AC}$ \\
\bottomrule
\end{tabular}
}
\end{table}
In Table 1, the $OP_{\text{MLP}}$ refers to the number of operations in the ChannelMLP, $OP_{\text{MLP}}=2D\cdot d_h$ for MLP with input dimension of $D$ and hidden dimension $d_h$.

\subsection{Confidence interval of the speaker identification task}
In this section, we evaluate the speaker identification task across 5 independent runs with different random seeds to ensure statistical reliability, as shown in Table~\ref{tab:Sid_1}. All results are reported as mean $\pm$ standard deviation. The Spiking-LEAF results are taken directly from the original literature. As demonstrated in Table~\ref{tab:Sid_1}, our IML-Spikeformer consistently achieves performance comparable to or better than the Transformer baseline while consuming significantly less energy, demonstrating the superior stability and efficiency of our IML-Spikeformer.

\begin{table}
\begin{center}
\caption{Results on VoxCeleb1 speaker identification task of 5 runs with random seeds.}
\label{tab:Sid_1}

    \begin{tabular}{cccccc}
    \toprule
    Model &Params& SNN& Timestep & Acc.(\%) &Energy(mJ)\\
    \midrule
    Spiking-LEAF& 0.9M& \CheckmarkBold & 1 & 30.45& 0.072\\
    \midrule
    \multirow{3}{*}{Transformer*}& 0.6M & \XSolidBrush & 1 &  67.21$\pm$0.33& 0.65$\pm$0 \\
                                & 1.18M & \XSolidBrush & 1 & 70.80$\pm$0.42& 1.03$\pm$0  \\
                                & 1.76M & \XSolidBrush & 1 &  72.81$\pm$0.25& 1.37$\pm$0\\
    \midrule
    \multirow{3}{*}{\textbf{IML-Spikeformer}}& 0.6M& \CheckmarkBold & 6 &  67.11$\pm$0.38& 0.16$\pm$0.05 \\
                                & 1.18M& \CheckmarkBold & 6 & 71.53$\pm$0.28 & 0.22$\pm$0.06\\
                                & 1.76M& \CheckmarkBold & 6 &  73.84$\pm$0.23 & 0.30$\pm$0.09\\
    \midrule
    \multicolumn{6}{l}{* \quad Our reproduced results based on publicly available codebases.} \\
    \end{tabular}
\end{center}
\end{table}
\subsection{Training convergence}
\begin{figure}[!t]
\centering
\includegraphics[scale=0.50,trim= 0 3 5 0, clip]{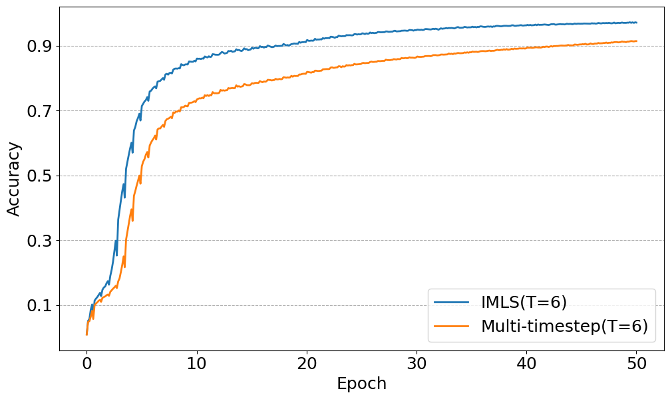}
\caption{Learning curves comparing IML-Spikeformer with iterative multi-timestep firing and our IMLS firing mechanisms. The curves demonstrate comparable convergence speed while IMLS achieves better final performance.}
\label{Fig: learning}
\end{figure}

In this section we provide the convergence speed comparison between the our IML-spikeformer with IMLS and the multi-timestep firing in Fig. \ref{Fig: learning}. This figure shows that our model with IMLS can achieve better performance with comparable converge speed to their performance limits. 

\end{document}